\documentclass[12pt,tightenlines,eqsecnum,floats,showpacs,nofootinbib,amsmath,amssymb,a4paper]{revtex4}

\usepackage{amsfonts,amsthm,amscd}
\usepackage{enumerate}
\usepackage[hang, flushmargin]{footmisc}
\usepackage{natbib}
\usepackage{enumerate}
\usepackage{graphicx}
\usepackage{hyperref}
\def\be{\begin{equation}}
\def\ee{\end{equation}}
\def\ba{\begin{eqnarray}}
\def\ea{\end{eqnarray}}

\def\L{\mathcal{L}}

\def\f{\frac}

\begin{document}

\title{$w=-1$ as an Attractor}
\author{David Sloan$^{1}$}
\email{david.sloan@physics.ox.ac.uk}
\affiliation{$^{1}$Beecroft Institute of Particle Astrophysics and Cosmology, Department of Physics,
University of Oxford, Denys Wilkinson Building, 1 Keble Road, Oxford OX1 3RH, UK}

\begin{abstract}

It has recently been shown, in flat Robertson-Walker geometries, that the dynamics of gravitational actions which are minimally coupled to matter fields leads to the appearance of ``attractors" - sets of physical observables on which phase space measures become peaked. These attractors will be examined in the context of inhomogeneous perturbations about the FRW background and in the context of anisotropic Bianchi I systems. We show that maximally expanding solutions are generically attractors, i.e. any measure based on phase-space observables becomes sharply peaked about those solutions which have $P=-\rho$. 

\end{abstract}

\maketitle

\section{Introduction}

The oft discussed measure problem in cosmology \cite{Gibbons:1986xk,Coule:1994gd} arises as a result of a symmetry between dynamical solutions to the equations of motion. In previous work it has been shown that the non-compactness symmetry group provides an explanation for the existence of attractors \cite{Corichi:2013kua}. This symmetry can be seen as the freedom to rescale the fiducial cell used in forming the finite dimensional Lagrangian used in cosmology from the field theory of gravity. In evaluating any measure on phase space, cut-offs must be imposed on the gauge direction, but these are not preserved under evolution. Thus, in accordance with Liouville's theorem, a spread in the evolution of one phase-space variable must be compensated by a focusing in others such that the total phase-space volume of a set of solutions remains fixed under the action of the Hamiltonian \cite{Sloan:2015bha}.

The Liouville measure is used to provide a notion of the probability of events occurring in a given dynamical system, with probability being defined as the relative volume in phase space \cite{Hawking:1987bi,Gibbons:2006pa}. Of course, such a measure can only give a ``raw" probability, as a measure can be used in conjunction with a variety of functions on phase-space variables to give different probability values. However, it is often argued that one should use the principle of indifference to argue that the raw definition is useful, and that events with either high or low raw probability would require functions with high information content to qualitatively change the result. This has been of particular interest when related to inflationary cosmology in which it was found that the probability one obtains is either very low  \cite{Gibbons:2006pa} or very high \cite{Measure} depending on the energy density at which the measure is based.

The existence of attractors explains the apparent incompatibility between results obtained at high and low energy densities \cite{Gibbons:2006pa,Page:2011yd,Measure2}. As was observed in \cite{Corichi:2010zp,Measure2} and replicated in \cite{Wald} these differences do not contradict Liouville's theorem. It turns out that they are in fact a direct consequence of it. The explanation of this in terms of attractors on inflationary phase space was provided in \cite{Corichi:2013kua} and this result was expanded to a wide range of physical systems and gravitational theories in the context of flat Robertson-Walker geometries \cite{MinCoup}. 

The purpose of this paper is twofold: Primarily we will explicitly derive the measure and attractors which are encountered in the context of perturbations on flat ($k=0$) Friedmann-Robertson-Walker models and show that by a choice of parametrization it becomes clear that any late time measure must be sharply peaked around maximally expanding solutions, i.e. those with $w=-1$. Secondly we will extend results to the anisotropic Bianchi I models, including anisotropic matter sources to show that isotropic, maximally expanding solutions are the attractors of the system. Throughout this paper we will use a scalar field to play the role of matter. It should be emphasised that this is done purely to give concrete examples of the attractor phenomenon, and our results apply in the case of any minimally coupled matter fields. The paper is laid out so that those familiar with the issues can use individual sections which can be read independently. In the following section we begin with a discussion of the symmetries of cosmological solutions under rescaling of the scale factor. Then in section \ref{Toy} an illustrative toy model is presented in which most of the analysis that will be used in the case of General Relativity (GR) can be seen in a simpler context. In section \ref{GR} we present the necessary formulation of GR for our analysis. In sections \ref{Background} and \ref{Perturbations} we present the behaviour of the background Friedmann-Lem\^aitre-Robertson-Walker model and the space of perturbations around it, and the global set of attractors for this setup follows in section \ref{Attractors}. The independence of the existence of attractors for the specific background model is shown in section \ref{Generic}, and finally in sections \ref{Discussion} and \ref{Conclusions} we present some discussion of related issues for measures (particularly the `Q-catastrophe' and eternal inflation), and the conclusions. 

\section{A Note on Symmetries} 
\label{Symmetries}

Before we begin our main analysis, let us first recall some elementary facts about cosmology and measurements which despite their simple nature appear to have been misunderstood in recent literature. The most substantial of these is that the scale factor, $a$ cannot be measured independently of some other length scale. In particular, when dealing with a homogeneous cosmology, there is no natural choice of length scale, as to form a length one would need to identify two separate points, which in turn would require that the points were distinguishable, breaking the assumption of homogeneity. If the universe is closed, an observer could consider sending a photon out and awaiting its return, using the time of travel to establish the circumference of the compact space. However, this would only determine the circumference up to a choice of time scale, returning the same issue. Using such a technique an observer could determine the relative anisotropy of the universe, by e.g. the number of times a photon orbits in one direction during the orbit of a photon in another, but still the overall length scale would remain undetermined. 

One might think that inhomogeneities would solve the issue. However, to dispel this idea, let us use a thought experiment: Consider a box of edge length $L$ inhabited by a field obeying the Klein-Gordon equation, subject to periodic boundary conditions. An observer given this box could perform a Fourier decomposition of the field and establish that $L$ is the wavelength $\lambda$ of the lowest order mode. However, once again this has only established $L$ in terms of other observables and not outright: Given a second box of edge length $2L$ the observer would have arrived at the same conclusion. Again the observer could establish differences in edge lengths of an asymmetrical box, but would only be able to determine the ratio of these as the ratio of lowest order modes. The volume of the box would be inaccessible. The problem is not resolved if we consider a different topology: Consider the same system but restricted to the surface of a sphere: The observer can once again establish the lowest order mode in a decomposition of the field into spherical harmonics. However the curvature of the sphere can only be determined in terms of this longest wavelength (indeed, the radius of curvature will always be $2\pi/\lambda$. Thus an observer with access to only one box cannot distinguish whether they are in box 1 or 2, which is the situation in cosmology: All measurements must be made from within the system. This should come as no surprise to readers familiar with a `rods and clocks' description of observables, or relational observables. The principle is simple: When specifying the length of on object we must give a reference length, such as the m\`etre des Archives in Paris. Under a change of base length to, say, inches, there will be an equivalent description; values of parameters will change but physics will not. Indeed it is interesting to note that from 1960 to 1975 the metre was defined to be equal to 1 650 763.73 wavelengths in vacuum of the radiation corresponding to the transition between the levels $2p_{10}$ and $5d_{5}$ of krypton-86 \cite{Baird:63}. 

An objection raised by a referee is that although there is no preferred scale in the $k=0$ cosmology, one could use the size of the universe at maximal extent ($\dot{a}=0$) to give a size which could be measured in terms of the radius of a hydrogen atom, distinguishing between universes that would otherwise be identified under the proposed symmetry relation. This example  is particularly subtle as the width of a hydrogen atom is determined in terms of a quantum mechanics, and the analysis presented here is entirely classical. Therefore to avoid confusion, let us substitute the (physical) metre. Again, the maximal radius of the universe could be measured in terms of this object, but a separate universe in which both the radius and the length of the metre were halved would be indistinguishable classically. This is key to the classical behaviour of the Liouville measure - the measure counts all sizes of Metre in Paris and equivalent rescalings of the maximal radius of the universe, size of the galaxy, etc, separately. However, to an observer these are indistinguishable. A universe half the size, with a milky way half the size, containing a half-sized earth, half sized paris and thus half size metre (size meaning linear length here - areas and volumes, momenta, time, scaled accordingly) would be considered a separate solution to the equations of motion, and therefore counted as a separate system by the Liouville measure. However, to the (half sized) observers in this universe, physics would proceed exactly as if the universe had the original full size. 

As we have noted, this analysis is entirely classical, and therefore does not directly deal with the objection that scale is inherent to quantum mechanical systems - the Bohr radius being a function of the Planck constant, speed of light, mass of the electron and the fine structure constant. We stress that the analysis we present is classical, and rests on theorems of classical Hamiltonian systems. However, the symmetry that we note can be seen to persist in the quantum regime in a more general theory space - if we consider universes in which not only scales vary but also, say, the values of dimensionful quantities in the standard model (such as the electron mass) there will again be multiple solutions that are indistinguishable to an observer, related by appropriate rescalings of lengths and (for example) the mass of the electron. Thus any measure on a larger theory space should take care not to count such solutions as separate entities but rather to identify such cases as providing the same observations to an observer within the system. 

Therefore, when we consider physical observables of our theory, we should always be cognisant that this choice of volume at any given time is pure gauge: under a rescaling $a \rightarrow \lambda a$ there will always be a choice of parameters which would give an identical set of physical observations. In homogeneous cosmology the symmetry is made apparent on writing the Friedmann equation in terms of the energy densities of perfect fluids:
\be H^2 = H_o^2 (\f{\Omega_m}{a^3} + \f{\Omega_r}{a^4} +\f{\Omega_k}{a^2} + \Omega_\lambda) \ee
It is trivial to see that there is a symmetry between solutions under 
\be \{a,\Omega_m, \Omega_r, \Omega_l, \Omega_\lambda\} \rightarrow \{\mu a, \mu^3 \Omega_m, \mu^4 \Omega_r, \mu^2 \Omega_k, \Omega_\lambda\}. \ee
This symmetry is normally fixed by setting $a_o=1$, but this is merely a choice of convention, much akin to the rescaling of the curvature of homogeneous, isotropic spatial manifolds such that $k= \pm 1$ in open and closed universes.   

When dealing with such systems we are not interested in the entire set of trajectories which can exist, but rather the set of distinct physical possibilities, where we identify two solutions if they are physically indistinguishable. Such an identification can be made through gauge fixing. We note here that a similar notion of relationalism underlies the theory of Shape Dynamics \cite{Gomes:2010fh, Barbour:2013jya}, and we should expect attractors to be universal in this context.

\section{Toy Model} 
\label{Toy}

In order to ease understanding of the analysis that will follow, let us first consider an illustrative toy model which exhibits qualitatively similar behaviour to that of our cosmological system. As we shall see, it is neither the specific form of the action under consideration, nor any initial conditions that give rise to attractor behaviour - rather it is the form of the coupling between fields. In the case of homogeneous, isotropic Roberston-Walker cosmologies, this was discussed in \cite{MinCoup} in which it was shown that there exist attractors for $F(R)$ theories for a large class of matter sources. Here we will examine first a simple Lagrangian system in which the attractor behaviour appears, then generalise this to include a more varied range of matter sources and kinetic terms. 

Let us consider a simple, single particle system defined by the Lagrangian
\be \label{Ltoy} \L = \f{\dot{x}^2 e^x}{2} \ee
this model is motivated by GR in which this would form the part of the Lagrangian associated with the expansion of scale factor in a cosmological model under the transformation of variables $x=\log(a)$. However, it will suffice for our present analysis to think of this simply as a model for the behaviour of a single particle. One can trivially obtain the momentum conjugate to $x$ as $P=\dot{x}e^x$, and hence the Hamiltonian is identical to the Lagrangian. Further, upon obtaining the equations of motion, we find that:
\be x(t) = x_o + 2 \log(t-t_o) \ee
and hence we find that 
\be \dot{x} = \f{2}{t-t_o} \quad P=2(t-t_0)e^{x_o} \ee
Let us suppose that we are interested in the distribution of $\dot{x}$, and its behaviour over time. We note immediately that the evolution of $\dot{x}$ is independent of a given value of $x$ at any initial time, that is to say that there exists a dynamical similarity under $x \rightarrow x+\lambda$ for any real $\lambda$, which is obtained simply by reassigning the value of $x_o$. We can therefore choose an arbitrary value of $x_o$ for any given trajectory which will be irrelevant to its behaviour, and we see the dynamics of $\dot{x}$ are entirely determined by the value $t_o$. Mathematically, we can consider an equivalence relation between solutions to the equations of motion, $\sim$ defined by

\be S^a \sim S^b \leftrightarrow \dot{x}^a = \dot{x}^b \ee
for any two solutions $S^a$ and $S^b$ wherein $\dot{x}$ is evaluated at a given time $t$. Thus a choice of a representative of each equivalence class is a choice of $x_o$. For distributions of $\dot{x}$ we are therefore only interested in the space of solutions modulo this equivalence. 

Consider a set of trajectories uniformly distributed at $t=1$ between $\dot{x}=1$ and $\dot{x}=2$. For a selected trajectory at this time, we would state that the probability that the trajectory had $\dot{x}$ under some value would be given:
\be P(\dot{x}<U) = \int_1^U du = U-1\ee
a simple uniform distribution on the interval $[1,2]$. However, at a later time, $t=2$ say, these trajectories will no longer be uniformly distributed over an interval. In order to count the same set of trajectories, we would need to evaluate their relative positions at $t=2$. Using a subcript to denote the time at which the value is observed, we find that:
\be \dot{x}_1 = \f{2 \dot{x}_2}{2-\dot{x}_2} \rightarrow \f{d\dot{x}_1}{d\dot{x}_2} = \f{4}{(\dot{x}_2-2)^2} \ee
and hence to obtain the same distribution we would no longer use a uniform distribution over $\dot{x}_2$ but would rather have to replace the differential element appropriately:
\be P(\dot{x}_1<U) = \int_{2/3}^{U'}  \f{4}{(\dot{x}_2-2)^2} d\dot{x}_2 \label{Toyprob} \ee
wherein $U'=\f{2U}{2+U}$ is the value that the trajectory passing through $\dot{x}_1=U$ takes at $t=2$, and the trajectory passing through $\dot{x}=1$ at $t=1$ reaches $\dot{x}_2 = 2/3$. Thus we observe that the uniform distribution is no longer valid, but rather a re-weighting due to the focussing of some trajectories has occurred.  What was a uniform distribution at one time becomes dependent on the parameter at a later time.

This behaviour is unsurprising - there is no reason why under evolution a set of trajectories spread across some parameter should retain the form of their distribution. A set of equal length pendulums set swinging with differing amplitudes will see their relative displacements shrink and grow throughout their oscillations. However, when one examines behaviour of both position and momentum, Liouville's theorem states that the distributions in phase-space are preserved: A narrowing of the spread of positions of our pendulums is compensated by an expansion in the spread of their velocities. 

The same system can be considered using Liouville's theorem: In this case the symplectic structure $\omega$ is $dx \wedge dP = \dot{x} e^x dx \wedge d\dot{x}$. Under dynamical evolution, the phase-space area as measured by this two-form will be conserved, and thus we can see how this can be used to induce a probability measure at time $t=2$ from a uniform distribution at $t=1$ in the same manner as was done above. First we note that the uniform distribution on $\dot{x}$ can be thickened to a measure on phase space by noting that at $t=1$:
\be P(\dot{x}<U) = \int_1^U d\dot{x} = \f{1}{N} \int_1^U \int_I \f{\omega}{\dot{x}} \label{Sympform} \ee
in which $I$ is the interval over which we have thickened the measure, and $N$ is the total phase-space area measured by $\omega$. This is a constant under evolution, and serves to normalise the total probability to unity. For the equality to hold, the interval $I$ is chosen to be uniform on $\dot{x}$ at $t=1$. The initial choice of interval is arbitrary, and we shall therefore choose it such that it has unit length as measured by the projection of the symplectic structure into the $x$ direction - i.e. 
\be \int_I e^x dx = \int_I d(e^x) = 1 \ee
Such an interval is given between limits $e^x \in [1,2]$, so $I$ is in the interval $x \in [0,\log(2)]$. 

\begin{figure}[h]
	\includegraphics[width=0.95\linewidth]{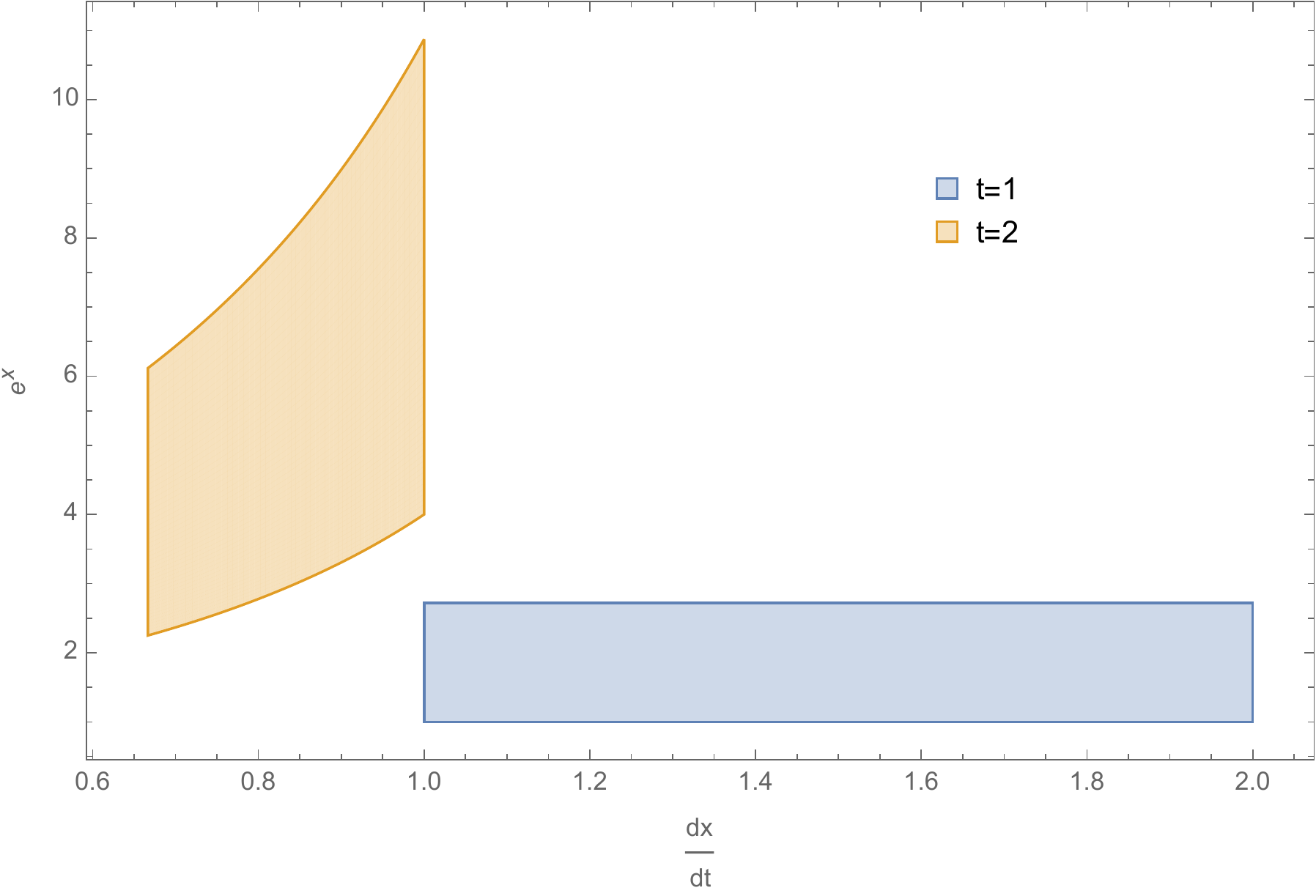}
	\caption{The space in the $\dot{x} - x$ plane occupied by our set of solutions at the initial time $t=1$ (blue) and later time $t=2$ (yellow). We see that the phase space area has been conserved, as per Liouville's theorem, but the shape occupied has changed. It is from this relative stretching that we recover the probability weighting of solutions spread in $x$.}
	\label{ToyPhaseSpace}
\end{figure}

\begin{figure}[h]
	\includegraphics[width=0.95\linewidth]{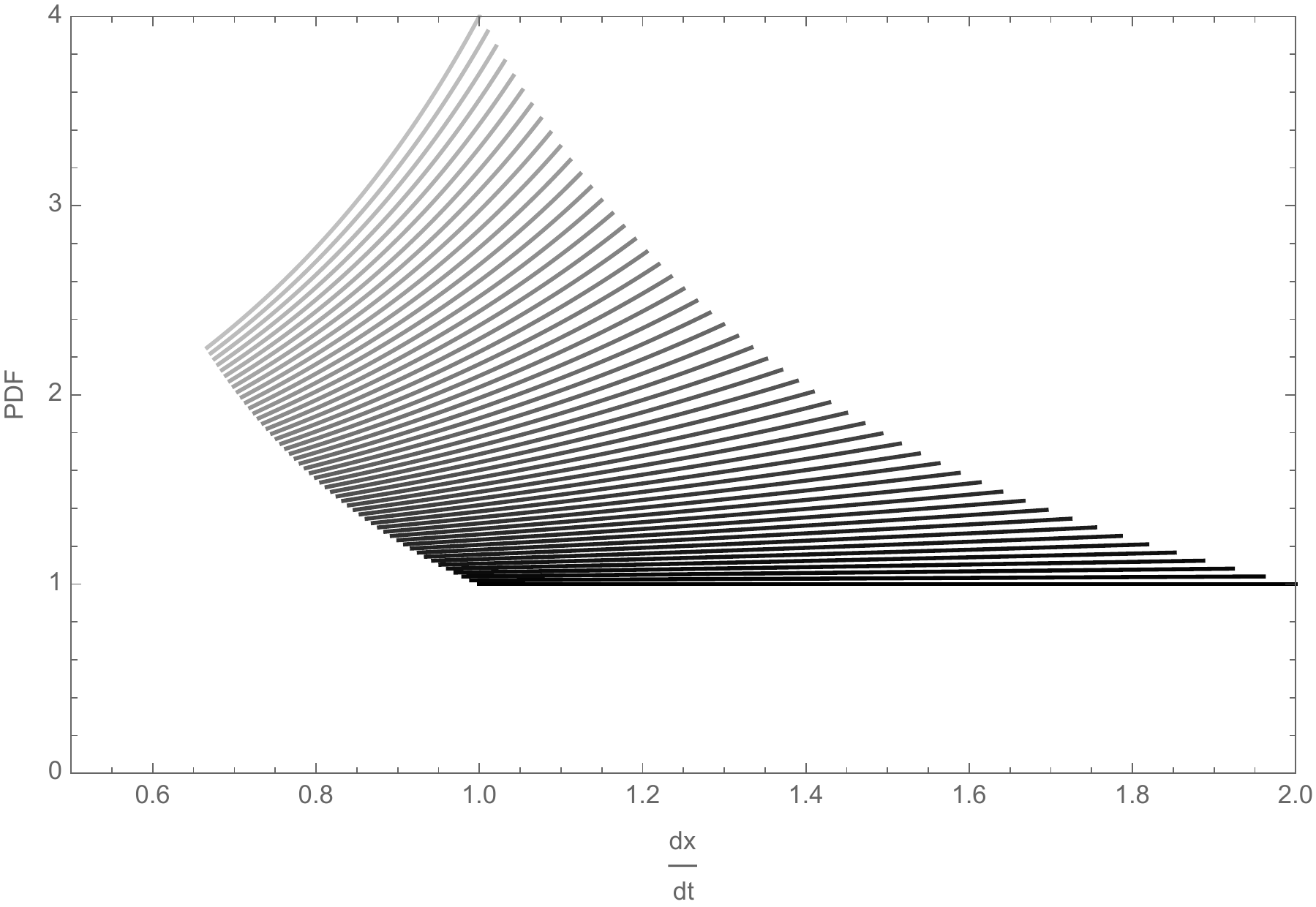}
	\caption{The probability density function for our toy model varies over time (shades of grey) showing the evolution of a uniform distribution at $t=1$ (black) to a re-weighted distribution at $t=2$, following the shape of the expansion of the interval in $e^x$ seen in the phase space picture.}
	\label{ToyProbDensity}
\end{figure}

Consider now the same set of trajectories at $t=2$. As we have seen above, the range over $\dot{x}$ in which these trajectories lie is $[2/3, U]$ (note that the dynamics of $\dot{x}$ is independent of the choice of $x$, and hence this interval is valid for any such choice.) The interval $I$ has also changed - not only has it moved, but it has been stretched, and the stretching is dependent on the value of $\dot{x}$: following the dynamics, the interval is now $x \in [-2\log(\dot{x}/2-1), 2\log(2)-2\log(\dot{x}/2-1)]$. This is demonstrated in figure \ref{ToyPhaseSpace} Thus, when we perform the integral over $x$ to recover a probability distribution on $\dot{x}$, projecting our measure on phase-space back onto one simply on the parameter of interest, we find that the integral over the extra direction yields:
\be \int_I e^x dx = \int_{\f{1}{(\dot{x}/2-1)^2}}^{\f{2}{(\dot{x}/2-1)^2}} d(e^x) =\f{4}{(\dot{x}-2)^2} \ee
Thus we recover exactly the same probability weighting. We can then project this measure back down onto the range of $\dot{x}$ defined using equation \ref{Sympform} to find:
\be (\dot{x}<U) = \int_{2/3}^{U'} \f{4}{(\dot{x}-2)^2} d\dot{x} \ee
This is exactly the same as was found by solving the equations of motion for $\dot{x}$. We can generalise this across time noting that all that is required is the form of $x(t)$ for any given later time $t$. Doing so we find that the relative weighting $W(\dot{x})$ becomes
\be W(\dot{x}) = \f{4}{(2+(1-t)\dot{x})^2} \ee
noting that this weighting returns to a uniform distribution at $t=1$. This is show in figure \ref{ToyProbDensity} in which we see the evolution of the probability density function between the uniform distribution at $t=1$ and the weighting encountered at $t=2$. Further we should note that the weighting was independent of the choice of initial distribution of probability: We had chosen for simplicity the uniform distribution, however for any probability density function $f(\dot{x})$  we could have performed the exact same procedure;
\be P(\dot{x} < U) = \int_1^U f (\dot{x}) d\dot{x} = \f{1}{N} \int_1^U  f(\dot{x}) \f{\omega}{\dot{x}} \ee

and under evolution the behaviour of the interval $I$ in $x$ is unaffected by the choice of probability density function. Therefore given \textit{any} probability measure at one time, we can calculate the equivalent measure at a later time, and find it always to be a re-weighting by expansion in the perpendicular parameter.  As usual, the distribution itself must be altered to follow the evolution, but the relative weighting is independent. 

 However, in this case we did not need to evaluate the probability based on the evolution of $\dot{x}$ but rather we utilised the conservation of the symplectic form under evolution that is given by Liouville's theorem. What is important to notice is that it is not the precise form of the integral over $x$ that was important, but rather the fact that the interval over which the integral was performed underwent stretching, and that the stretching was dependent on the initial parameters. Upon projection back onto the physical parameter of interest, this stretching has become a probability weighting, weighting most heavily on those solutions which underwent the greatest expansion in the $x$ direction. Even though the value of $x$ at any given time was arbitrary (this constituted the choice of representative member of the equivalence classes of solutions under $\sim$), the relative dynamics of $x$ for each solution allows us to recover the distributions of $\dot{x}$ at different times. 

We see therefore that Liouville's theorem has given us a powerful tool. It allows us to induce a probability measure on a physical parameter at a given time from a measure at an earlier time and the dynamics of \textit{other} fields. Those solutions for which the relative weighting becomes highest we will term `attractors', as they are trajectories through phase space about which the distribution on physical parameters of interest (in this case $\dot{x}$) become focused. When we come to cosmology, the particular direction of interest will be that represented by the volume of space. Although this is not an observable at any time, its relative expansion between times will allow for the attractor behaviour of other parameters to manifest.

\section{GR Action and Symplectic Structure} \label{GR}

We will examine inhomogeneities in the context of perturbations around a FRW background. For ease of exposition we will work with matter in the form of a massive scalar field, the most common inflationary model. This is not strictly necessary for the analysis that will be performed, and as will be shown analogous results can be obtained with any matter source.  We begin with the Lagrangian density for GR coupled to matter and follow the standard procedure as outlined in \cite{ABR}. Let $R$ be the Ricci scalar, then: 
\be \L = R - \L_{matter} \ee
The Einstein-Hilbert action is therefore given
\be S = \int_M \L \epsilon_{abcd} dV^{abcd} = \int_M (\sqrt{g}R - \sqrt{g} \L_{matter}) d^4 x \ee
In which we see the minimal coupling between gravity and matter appear. For simplicity we will take the manifold $M$ to allow a decomposition into compact spatial slices and a time direction: $M=\Sigma \times \mathbf{R}$.  The effect of this minimal coupling between any gravitational and matter theories in a homogeneous, isotropic context was examined in \cite{MinCoup}.  Varying this action yields the usual Einstein equations coupled to matter, plus a boundary term. Since the equations are globally hyperbolic the surfaces $\Sigma_{t}$ are Cauchy slices and hence we obtain the familiar
\be \delta S = \int_{M} (Equations of Motion) \delta (fields) + \int_{\partial M}  J \ee 
where $J$ is our candidate for a presymplectic current. In the case of the Einstein-Hilbert action coupled to a matter field $\psi$ on our manifold this becomes:
\be \delta S = \int_{M} (G^{ab} - \kappa T^{ab}) \delta g_{ab} +(EoM)\delta\psi + \int_{\Sigma_1 \cup \Sigma_2} \pi^{ab} \delta q_{ab}  + P_\psi \delta \psi\ee
in which $h_{ab}$ is the induced metric on the Cauchy slices $\pi^{ab}=\sqrt{h}(K^{ab} - Kh^{ab})$ is the usual ADM momentum expressed here in terms of the extrinsic curvature of the slice. Following \cite{ABR} we find our symplectic current by second variation of the action about a solution to the field equations. Since our action is minimally coupled, the standard symplectic structure on the matter fields will apply, rescaling the matter momentum with a factor of $\sqrt{q}$ \cite{MinCoup}. Therefore we shall concentrate here on the gravitational part alone, and note that the linearity of the system allows us to form the complete symplectic structure as the sum of gravitational and matter components. Let us define $\delta_1 = \{ \delta_1 h_{ab}, \delta_1 \pi^{ab}\}$ and  $\delta_2 = \{ \delta_2 h_{ab}, \delta_2 \pi^{ab}\}$. Thus our presymplectic current is given:
\be J (\gamma, \delta_1 , \delta_ 2) =  \delta_1 \pi^{ab} \delta_2 q_{ab} - \delta_2 \pi^{ab} \delta_1 q_ab \ee
As we have seen this is an exact form, and is therefore closed hence 
\be \int_M dJ = \int_{\Sigma_2} J - \int_{\Sigma_1} J = 0 \ee
Hence $J$ is conserved between Cauchy slices. Hence we can integrate the presymplectic current over a Cauchy slice to obtain the presymplectic structure
\be \omega(\delta_1, \delta_2) = \int_\Sigma \delta_{[1} \pi^{ab} - \delta_{2]} q_{ab} \label{SyS} \ee

To obtain a symplectic structure we must further mod out gauge directions. In order to achieve this in the case of interest to this paper we will examine perturbed Friedmann-Robertson-Walker geometries following \cite{Mukhanov} and \cite{Bardeen}. Our analysis will be similar to that performed in \cite{Wald}, however an extra gauge symmetry in reducing from phase-space variables to physical observables will render a different result. 

\section{Background model and Symmetries} \label{Background}

Let us consider a Friedmann-Robertson-Walker metric with linear perturbations, the standard model used to describe inflationary observations. The background model is treated in \cite{Corichi:2013kua,MinCoup}, the results of which we will briefly summarize here. It is shown that considering only the homogeneous modes, the phase space can be expressed in terms of $\nu$, the volume of a region measure according to a fiducial cell, and the scalar field $\phi$. The phase space is therefore $\{\nu, H ; \phi, \pi\}$ and one finds a symplectic structure:
\be \omega_o = d\nu \wedge dH + d\pi \wedge d\phi \ee
On restricting to solutions to the field equations (in this case the Friedmann equation) at a fixed value of the Hubble parameter $H$;
\be \overleftarrow{\omega_o} = \sqrt{H^2 - V[\phi]} d\nu \wedge d\phi \label{backgroundmeasure} \ee
which is conserved under the Hamiltonian flow that provides evolution between Hubble slices. The procedure for a general matter source is the same - we impose the Hamiltonian constraint and pull back the symplectic structure to a surface of constant curvature. Since we are dealing with a homogeneous background we can ignore spatial derivatives of our matter fields. Schematically we will then obtain a volume form from the symplectic structure by raising it to a high enough power to cover all of phase space:
\be \overleftarrow{\Omega_o} = \sqrt{H^2-V[\phi]-\rho_r} d\nu \wedge d\phi \wedge d\vec{f} \wedge d\vec{\pi_f} \ee
in which $\rho_r$ represents the energy density of the remaining fields, $\vec{f}$, and $\vec{\pi_f}$ their momenta. 

Since physical observables are independent of $\nu$, there exists a symmetry between solutions under $\nu \rightarrow \lambda \nu$ in which $H$, $\phi$ and $\dot{\phi}$ remain unchanged. This is unsurprising as we must measure $\nu$ against some arbitarily chosen fiducial cell, and altering this extraneous structure should not render any differences in our physical observables. Indeed our physical setup would have been ill-defined if this were not the case. Hence we must fix an interval in $\nu$ to project this measure back onto the space of observables. On doing so, the resulting measure is no longer preserved under time evolution, and we find that those solutions which undergo the most expansion are given the highest relative weighting between Hubble slices. This is the nature of the attractor solutions. A more sophisticated form of this reasoning is followed in shape dynamics\cite{Anderson:2002ey,Gomes:2010fh,Barbour:2013qb} wherein the basic entities are 3-geometries modulo conformal transformations - this has been shown to coincide with the ADM formulation of GR in a constant mean curvature slicing \cite{Koslowski:2013qt}. 

To make this more explicit, consider an interval in field parameter $\phi$ on some initial Hubble slice, and thicken this by expanding at each point by some arbitrary interval in $v$. Without loss of generality, we will consider a uniform width in the gauge direction on the initial slice. Then at some later slice, the measure of this area given by \ref{backgroundmeasure} will be unchanged. However along the new interval in $\phi$ the volume direction will have expanded by a factor dependent on the initial conditions:
 \ba \log[\f{\nu_f}{\nu_i}] &=& 3 \int_{H_i}^{H_f} H[\phi,\dot{\phi}] dt \label{Expansion} \\
 			   &=& \f{2}{3} \int_{H_i}^{H_f} \f{d \log H}{1+w[H]}  \ea
On projecting back down to the physical observables this relative expansion in $\nu$ can be considered to be an induced probability distribution on $\phi$, and from the above dynamics the highest weighting is given to those solutions that undergo the greatest expansion. This generalizes to any matter source \cite{MinCoup}, and hence for regular matter subject to the usual energy conditions, will give strongest weighting on those for which $P=-\rho$ for the largest interval. Hence $w=-1$ is an attractor of the background dynamics. 

Once we introduce perturbations into our model, it would appear that the symmetry of the system has been broken, as the size of our fiducial cell can now be determined from the wavelengths of perturbations within it, and thus $\nu$ would become a physical observable. However upon examination a symmetry remains: If we increase both the length of the edge of our fiducial cell and the wavelengths of all inhomogeneous modes within it by the same factor we recover a physically indistinguishable configuration. Thus if we consider inhomogeneities to have wavelengths $\vec{k_i}$ under the rescaling $\{\nu, \vec{k_i}\} \rightarrow \{\lambda \nu, \lambda^{1/3} \vec{k_i}\}$ we should expect physics to be invariant.

In practical terms we will only be interested in a finite number of wavemodes measured against this fiducial cell, thus in the flat case our system will be described by a set of Fourier modes. In general the modes will be eigenvectors of the Laplace operator compatible with the fiducial background spatial slice, which will become more important in the case of anisotropic backgrounds considered later. Note that since we are using the Laplace operator compatible with the fiducial cell, rescaling of the volume does not affect the space of eigenvectors.

\section{Perturbations about FRW} \label{Perturbations}

We shall take as a guide the dynamical system consisting of linear perturbations about the FRW background described above. This will be used purely for the purposes of illustrating our point; as we shall argue later the results we obtain will be independent of the precise nature of the system under consideration. The space of linear perturbations about a homogeneous background was described in \cite{Bardeen,Mukhanov}. These perturbations can be split into their scalar, vector and tensor components. We will consider only a scalar field as our matter contribution to keep the description simple. Extension to a variety of matter sources does not qualitatively change the results obtained. Since we consider only a scalar field, vector perturbations to our background geometry integrate to zero, and hence we are left with only tensor and scalar modes. The tensor modes are the simplest to deal with: Let $t_{ab}$ be a traceless, symmetric, divergence free perturbation to the background metric. We can express each perturbation at a point in terms of a basis $h^i_{ab}$ at each point on our Cauchy slice. Note that since the spatial manifold is homogeneous, the space of perturbations is independent of the point on the manifold at which the perturbation is made. We can therefore expand a perturbation on the entire spatial section as
\be t_{ab}(\vec{x}) = \sum a_{ij} h^{i}_{ab} f_j(\vec{x}) \ee
In which $i$ runs over the basis of perturbations (i.e. from 1 to 2 to represent the two polarizations, $h_+$ and $h_\times$ ) and $f_j(\vec{x})$ is the set of eigenfunctions of the Laplace operator compatible with the background. In the case under consideration here, this is just the space of Fourier modes of our fiducial cell. Noting then that these functions form an orthonormal basis we can apply our definitions of the symplectic structure \ref{SyS} to find:
\be \omega_{T} = \nu \sum_{ij} d\dot{a_{ij}} \wedge da_{ij} \ee
Furthermore since these perturbations are compatible with the field equations (they represent free gravitational waves whose presence is a symmetry of the Ricci scalar) we do not need to pull this back to the space of solutions to the field equations. 

The treatment of scalar perturbations follows a similar pattern. There are five functions which define a scalar perturbation, however there is a gauge symmetry which removes two of these and following \cite{Bardeen} we deal only with gauge invariant quantities, the Bardeen potentials. As above, we consider the space of perturbations expanded in Fourier modes, $f_j(\vec{x})$: $\Phi(x) = \sum_j b_{j} f_j(\vec{x})$ Following \cite{Wald} we impose the linearized field equations and ignore terms that vanish on integration over a Cauchy slice, to obtain
\be \overleftarrow{\omega_{s}} = \nu \sum_{j} d\dot{b_j} \wedge db_j\ee
after some normalization of the amplitudes $b$. Thus we obtain the total symplectic structure of our system, 
\be \overleftarrow{\omega} = \label{TotalSyS} \overleftarrow{\omega_o}+\overleftarrow{\omega_s} + \overleftarrow{\omega_T} = \sqrt{H^2 - V[\phi]} d\nu \wedge d\phi + \nu \sum_{ij} d\dot{a_{ij}} \wedge da_{ij} + \nu \sum_j d\dot{b_j} \wedge db_j \ee
Again it is important to note that had we performed the same analysis with any matter source we would have obtained an analogous result - by expressing each field in terms of Fourier modes and enforcing the equations of motion we would once again find the total symplectic structure expressed in terms of the amplitudes of Fourier modes for each gauge invariant quantity.\footnote{One notable difference would be the inclusion of vector modes, which would add a fourth component to the symplectic structure. This would, however, take an analogous form to those already presented.} On doing so, regardless of the matter model used, so long as the only interaction with gravitational degrees of freedom is achieved through minimal coupling, one obtains a symplectic structure of the form:
\be \overleftarrow{\omega} = \sqrt{H^2 - V[\phi]-\rho_r} d\nu \wedge d\phi + \nu \sum_j dm_j [f] \wedge d\pi_m^j [f] \ee
in which the gauge invariant fields, $f$ and their momenta $\phi_f$ are expressed in terms of their Fourier modes $m_j[f]$. The scalar field itself is not strictly necessary at this point: One only needs a background mode to eliminate in the homogeneous component of the Hamiltonian constraint so that the symplectic structure can be pulled back on to the solution surface. 

\section{Attractors} \label{Attractors}

We will now demonstrate the existence of attractors on the space of physical observables of our system. As discussed above, we will truncate to a finite number $l$ of Fourier modes on space, thus the dimension of phase space at a fixed value of the Hubble parameter is $N=6l+2$ arising from 2 polarizations for each tensor mode, 1 scalar mode, the background mode of the scalar field, the volume, $\nu$ of our space and associated conjugate momenta, modulo the our constraints.  We form a measure on this space by raising \ref{TotalSyS} to a volume form:
\be \Omega = \overleftarrow{\omega}^{3l+1} =\sqrt{H^2-V[\phi]} v^{3l} d\nu \wedge d\phi \bigwedge_{i=1,2 j=1}^{j=l} d\dot{a_{ij}} \wedge da_{ij} \bigwedge_{j=1}^l d\dot{b_j} \wedge db_j \ee
By Liouville's theorem, this volume form is independent of the choice of Hubble parameter, $H$, and thus any set of solutions measured according to this at some initial value, $H_o$ will have an equal volume when measured at $H_f$. In fact, since $H$ is a monotonic non-increasing function in General Relativity, we can use this measure to count the total number of solutions to the equations of motion.  Since the volume parameter is unbounded, taking values in $\mathbf{R_+}$, this total measure is infinite. However, we are interested only in physically distinct solutions, and since we are free to rescale the volume we will be considering measures which are independent of this gauge parameter.  Let us consider a general measure of this type: We will require a $(6l+3)$-form to cover the space of physical observables of our system, which we will pull back to the intersection of the constraint surface and $H=constant$. Therefore we obtain at $(6l+1)$-form:
\be M_H =F_H [\phi, a_{ij}, \dot{a_{ij}}, b_j, \dot{b_j}] d\phi \bigwedge_{i=1,2 j=1}^{j=l} d\dot{a_{ij}} \wedge da_{ij} \bigwedge_{j=1}^l d\dot{b_j} \wedge db_j \ee
Note that we associate the subscript $H$ with the function $F$ to state that this should be defined on some given Hubble slice. $F$ can be independent of the choice of slice, but we leave open the possibility that $F$ is slice dependent. We can then use $F$ to determine a probability distribution by defining the probability of an event to be the integral over physical parameters for which the event happens, normalized by the total integral of $F$ over all values of our physical parameters. Any probability distribution must take this form for some function $F$. 

We can now ``thicken'' our measure $M$ by taking its wedge product with $d\nu$, and considering that all physical observables are independent of the choice of $\nu$ we obtain the same probabilities of events by integrating this thickened measure over some arbitrary finite interval in $\nu$.\footnote{This is akin to taking any arbitrary probability distribution in $n$ variables and extending it over a (compact) extra dimension, then integrating out that extra dimension. It is trivial to show that this does not affect the probabilities of events occurring} In doing so we obtain a volume form which, by uniqueness, must be proportional to $\Omega$: 
\be T_H = M_H \wedge d\nu = \f{F_H[\phi, a_{ij}, \dot{a_{ij}}, b_j, \dot{b_j}]}{\sqrt{H^2 - V[\phi]}  \nu^{3l}} \Omega \ee
in which we introduce the ratio of $F_H$ and $\sqrt{H^2 - V[\phi]}$. Thus, by the inverse of this process (integrating $T_H$ over some interval in $\nu$ and dividing out by the length of the interval) we can recover any probability distribution on the space of physical observables. 

The Liouville measure $\Omega$ is conserved between Hubble slices. However, any interval in $d\nu$ over which it is integrated is not. Thus we can follow the analysis of \cite{Corichi:2013kua} and interpret this relative expansion as an induced weighting: When measured at $H_f$, trajectories as measured at $H_i$ are weighted by the relative increase of the length of the interval in $\nu$ between $H_i$ and $H_f$. Since the total volume of the $(6l+2)$-form $\Omega$ is constant, an expansion in the $\nu$ interval must be compensated by a contraction on the remaining (physically observable) phase-space variables.  To see this simply, consider two trajectories between these slices, $\gamma_1$ and $\gamma_2$ whose volumes expand by factors of $\lambda_1$ and $\lambda_2$ respectively between initial and final slices . On $H_i$ slice both have been thickened by the same interval $[\nu_s, \nu_e]$ in the volume direction. By $H_f$, $\gamma_1$ occupies the interval $[\lambda_1 \nu_s, \lambda_1 \nu_e]$ and $\gamma_2$ occupies the interval $[\lambda_2 \nu_s, \lambda_2 \nu_e]$, so to measure them with the same measure at $H_f$ as they had at $H_i$ we must integrate them over different ranges of $\nu$. However, since dynamics is independent of $\nu$ this can be achieved by integrating each over the same interval, and reweighting $\gamma_2$ by $\f{\lambda_2}{\lambda_1}$. Thus measuring the trajectories at $H_i$ instead of $H_f$ is equivalent to introducing a weighting proportional to the relative expansion in volume. Note that this stands in contrast to introducing volume-weighting by hand \cite{Linde:2007nm,Winitzki:2008yb,Hawking:2007vf} - the volume weighting is induced by the conservation of the Liouville measure. 

Since we have found that a gauge symmetry in our volume parameter $\nu$ induces a probability weighting on measures on physical observables, it is natural to investigate the behaviour of other symmetries of our system. Consider, for example, a scalar field $\phi$ for which the potential is either zero (i.e. the field is massless) or cyclic on some interval of length $L$. Then clearly there exists a gauge choice in the field parameter $\phi$ under $\phi \rightarrow \phi + \mu$, in which $mu$ is can take any real value in the case of a massless field, or any integer multiple of the interval $L$ for cyclic potentials. Then to generate a measure proportional to the Liouville measure from a measure on the space of gauge invariant observables, we must again either thicken our measure to by taking the wedge produce with $d\phi$ on the case of a massless field, or restrict our interval of integration over $\phi$ to being the interval length $L$\footnote{Strictly speaking, any integer multiple of the length $L$ could be considered, but this would render an indistinguishable result.} However, since $\dot{\phi}$ is independent of $\phi$ (or under $\phi \rightarrow \phi + nL$) the interval length over which integration is carried out will be unchanged on evolution between Cauchy slices, thus projecting down from a the Liouville measure to a measure on physical observables by integrating over a fixed interval in the gauge parameter $\phi$ is conserved by evolution. In the case of volume things are different because the gauge symmetry is not $\nu \rightarrow \nu+n$ but rather $\nu \rightarrow \lambda \nu$, and hence interval lengths are not invariant under the gauge transformation.

\section{Maximal Expansion is Generic} \label{Generic}

The model which we have used as illustration so far has not included any backreaction between perturbative modes and the background, nor any interaction between modes. We could expand beyond the linear regime following \cite{Uggla:2011jn}, however it is possible to consider the complete theory directly following the methods above and note that interactions between modes would have no qualitative effect on the analysis we have performed.  Consider a general action for gravity minimally coupled to matter fields, for which we will presume a constant mean curvature foliation exists\footnote{This is not highly restrictive - see \cite{Rendall} for example.} On decomposing into the background volume $\nu$ and remaining physical observables in terms of Fourier modes $\vec{m}$ the Liouville measure pulled back to intersection of the space of solutions with a constant Hubble slice is:
\be \Omega=f[H, \vec{m}] d\nu \wedge \Psi \ee
in which $\Psi$ is a volume form on the space of $(n-1)$ of the physical observables, with $f$ enforcing the Hamiltonian constraint through elimination of the final physical observable. If our matter satisfies the weak energy condition we are guaranteed that $H$ is monotonic, and thus we can use this measure to count solutions. This decomposition works in the case of background anisotropies also. Let us examine the question of why anisotropies tend to be suppressed\cite{Collins:1972tf}. Consider the line element for the Bianchi I model:
\be ds^2 = -dt^2 + \nu(t)^{2/3} \left( e^{-2\sigma}dx^2 + e^{\sigma-\f{\beta}{\sqrt{3}}} dy^2 + e^{\sigma+\f{\beta}{\sqrt{3}}}dz^2\right) \ee
Then we find the Ricci scalar is given:
\be R = \f{\ddot{\nu}}{\nu} + \left(\f{\dot{\nu}}{\nu}\right)^2 +\f{\dot{\sigma}^2}{2} + \f{\dot{\beta}^2}{2} \ee
Note that then in the action the second derivative term in $v$ becomes a total derivative, and hence yields only a topological contribution. 
\be S_g+S_m=\int \sqrt{-g}R + \sqrt{-g}\L_m = \int \nu \left(-\left(\f{\dot{\nu}}{\nu}\right)^2 + \f{\dot{\sigma}^2}{2} + \f{\dot{\beta}^2}{2}\right) + v\L_m \ee
Thus the anisotropies can be seen to have the same action as homogeneous massless scalar fields. We can trivially form the symplectic structure and hence the Liouville measure on such a system: The background portion $\omega_o$ when pulled back to a surface of constant Hubble curvature is now given:
\be \overleftarrow{\omega_o} = \f{2 H^2}{\sqrt{2 H^2-\dot{\beta^2}}} d\nu \wedge d\sigma \wedge d\dot{\beta} \wedge d\beta \ee 
Again, as expected the measure is composed of an integral over the gauge parameter $\nu$ and the physical degrees of freedom. Note that in the Bianchi I case $\sigma$ and $\beta$ can be shifted and the system is symmetric under $\{\beta, \sigma\} \rightarrow \{\beta + \lambda_1, \sigma +\lambda_2\}$ for arbitrary $\lambda_1 , \lambda_2 \in \mathcal{R}^2$. The lengths of any gauge interval in these parameters is maintained under evolution, therefore this gauge freedom will not affect results in the way that the freedom to rescale $\nu$ does. Furthermore,  the modes into which we decompose the inhomogeneous degrees of freedom should be compatible with the Laplace operator on the geometry determined by the spatial slices of this metric, which will introduce a coupling between the anisotropic shear terms and the matter degrees of freedom. We could consider a perturbative analysis in the mould of \cite{Pereira:2007yy}, and examine the space of linear perturbations as an illustrative guide. However even in the full non-perturbative dynamics there is no further coupling with $\nu$, and this is all that is required for our investigation. These anisotropies, though small, are potentially observable through weak lensing observations \cite{Pitrou:2015iya,Pereira:2015jya}.  The dynamics of this system have been studied in a the regime of quantum cosmology \cite{Gupt:2013swa}, establishing both the presence of the attractor and the resolution of singularities.  

We are now in a position to state the primary result of this analysis: Suppose that at some initial Hubble slice $H_o$, the state of the physical observables of our system is distributed according to some probability density function $T$. Then the relative probability of observing a set of trajectories with some physical property at a later Hubble slice $H_c$ is proportional to the relative expansion of the universe along each trajectory between $H_o$ and $H_c$. Those with the greatest expansion have the highest probability of being observed. In particular, those solutions which have $w=-1$ (i.e. $P=-\rho$) for the greatest duration will be those observed the most frequently. Furthermore, as we take the limit in which the initial slice is taken to the big bang, $H_o$ becomes infinite and the relative weighting becomes a delta function on those solutions - when measured at late times universes which experienced a great duration in which $P=-\rho$ will dominate completely outside a set of measure zero.

Let us again stress some of the things on which we have not relied: none of our analysis depended qualitatively on the physical situation under consideration here - an arbitrary matter Lagrangian with an arbitrary set of Fourier modes would produce the same result - the maximum allowable expansion will be that which is focussed upon. If $P=-\rho$ is permissible in a system, it will be achieved. Further there was no specific measure under consideration, as any can be thickened to a volume form proportional to the Liouville measure. At no point were any thermodynamical/statistical mechanical properties of the system required (e.g. ergodicity): We simply need that at some early Hubble slice the physical observables of our system are distributed according to some very generic probability distribution function. Our results can be overcome by specifying a very special distribution - one that has precisely zero support on trajectories that ever achieve $w=-1$ say - but this would require a great deal of fine-tuning to achieve. 

\section{Discussion} \label{Discussion}

As we have shown in the previous section, given a matter Lagrangian minimally coupled to gravity, solutions which undergo the most expansion in volume form attractors. The analysis that we present is independent of the choice of matter Lagrangian, being simply a consequence of the nature of the coupling and the conservation of the symplectic form due to Liouville's theorem. This in turn lead to a volume weighting of solutions. In the literature, such volume weighting of solutions in a multiverse context lead to a serious problem, known as the `Q-Catastrophe' \cite{Feldstein:2005bm,Garriga:2005ee}. We will sketch out the reasoning behind this problem in the context of a matter Lagrangian consisting of a massive scalar field. The problem persists in other cases, but this will suffice to illustrate the point. The maximal expansion that a solution can undergo between two Hubble slices is a function of the inflaton coupling $m$, and is achieved by the slowed possible roll down this potential. Thus the number of e-folds between these two slices seen in equation \ref{Expansion} is the integral of the Hubble rate with respect to time. We note from our equations of motion that $\dot{H}=4\pi\dot{\phi}^2$. Assuming slow roll \cite{Alho:2014fha}, we can relate $\phi$ and its derivative
 \be \dot{\phi}=\f{m^2 \phi}{3H} \rightarrow 4 \pi \dot{\phi}^2 =\f{4\pi}{9} \f{m^4 \phi^2}{H^2} = \f{m^2}{3} \ee
 in which we note that Friedmann's equation becomes $H^2 = \f{4 \pi}{3} m^2 \phi^2$ in the slow roll context. We are therefore lead to conclude that the number of e-folds between these two points is given
 \be N_e = 3\f{H_i^2-H_f^2}{m^2} \label{nofolds} \ee
 This is inversely proportional to the square of the mass of the inflaton, and therefore if, within our Lagrangian, there are different allowable masses for the inflaton, we find that the weighting will favour those with lower masses, as they provide greater expansion.  Therefore we should predict that if there are multiple possible inflationary mechanisms (multiple fields with differing masses, for example) the lower mass directions will dominate phase-space at late times as attractors. 

This analysis becomes problematic once perturbations are introduced in the manner of section \ref{Perturbations}. Such perturbations are the seeds of structure formation and observed in the cosmic microwave background, in which it is seen that the amplitude of these fluctuations, $Q=\f{\delta \rho}{\rho}$ if of the order of $10^{-5}$. It has been noted that in order for inhabitable galaxies to form, that this value should lie between $10^{-4}$ and $10^{-6}$. One can further calculate the amplitude of the perturbation wave modes as they re-enter the Hubble horizon. For a single field, this is found to be given $Q=\f{H^2}{2\pi\dot{\phi}}=\f{V^{3/2}}{V'} = \f{\sqrt{3 \pi}}{2} \f{H^2}{m}$. Thus changing the mass of the inflaton will alter the amplitude of fluctuations. We should, therefore expect to see a much lower inflaton mass, and hence larger fluctuations than are observed in our universe. This is merely a sketch of the full problem, which contains many more subtleties based on analysis of the distribution of perturbations, but the general idea is that the amplitude of fluctuations observed is not consistent with a maximal expansion, and hence our volume weighting should be considered invalid. 

One should note that throughout the arguments on the `Q-catastrophe', the parameter which is varied to allow for maximal expansion is the inflaton coupling, a part of the matter Lagrangian \cite{Aguirre:2006ak}. However, our analysis has been based on the phase-space of initial conditions subject to a given matter Lagrangian. It is not our goal to explain why a certain matter Lagrangian is that which we experience, nor do we claim that this could be done through a phase-space analysis. It is certainly plausible that if the true matter Lagrangian is more complex than that of the Standard Model, and one includes an entire string landscape, then our predictions would indicate that there should be a preference in the landscape for a low inflaton masses which produce greater expansion as these would indeed constitute landscape attractors. However, these are attractors of a classical field theory, not a quantum theory. Since any such behaviour should happen in the quantum regime, we cannot extend our analysis there. Rather what we have demonstrated is that given a matter source and gravitational action, maximally expanding solutions to that action are attractors within the classical regime. 

A similar issue arises when considering the possibility of eternal inflation. This is a regime in the early universe in which it is posited that the quantum fluctuations of the scalar field we use as an inflaton can cause the Hubble parameter to rise in some regions of space, as first order correction due to the quantum fluctuations of the field can overcome the classical dynamics. As such it is possible that some areas of space will exhibit and endless oscillating de-Sitter phase, wherein regions of space in which the Hubble parameter has increased expand faster than those in which it decreased, and hence the can dominate any volume of space at a given coordinate time. This can lead to a cyclic behaviour, eternal inflation, and the spawning of a number of `bubble universes' in which the initial conditions take different values. Since this process is endless, one encounters classical counting problems when attempting to place a measure on the possible outcomes \cite{DeSimone:2008if,Linde:2007nm,Freivogel:2011eg}. To give a sketch of the behaviour of eternal inflation, we consider the same massive scalar field as above. In such cases, for eternal inflation to persist we would require that the probability of the field value increasing during an e-fold be greater than $e^{-3} \sim 1/20$ \cite{Guth:2007ng}. To first order, the quantum spread of the field is approximately given:
\be \Delta \phi_q = \f{H^2}{2\pi} \ee
and so we find that in order for our behaviour to iterate, we would require that this perturbation outweigh the classical motion $\Delta \phi_c$ with probability approximately $1/20$. Hence we are lead to the condition, 
\be \f{H^2}{\dot{\phi}_c} > 3.8 \rightarrow \rho > m = 10^{-6} \rho_{pl} \ee
wherein we used the standard mass for the inflaton of $10^{-6}$ in Planck units. Therefore we would expect eternal inflation to persist at high energy densities, and spawn a number of bubble universes each behaving classically once the energy density drops below this level. 

Again, we make no claims as to the effects of eternal inflation, nor of its happening. It is tangential to the arguments that we put forth, as it may serve as a method of populating the ensemble of possible universes which then undergo a classical phase of expansion. This ensemble, however the prior distribution is determined by any pre-classical inflationary phase (be it eternal inflation, bouncing cosmologies or any other scheme) will then begin its classical phase of expansion in which any prior distribution of physical quantities at a high value of the Hubble parameter must be re-weighted for measurement at low energy density according to the attractor volume expansion. As a point of interest, we can use this to estimate the strength of the attractor between the end of eternal inflation and observations: The maximal expansion is given by equation \ref{nofolds}, with the choice of final Hubble parameter largely arbitrary as it is orders of magnitude lower than the onset. Thus such maximally expanding space-times will undergo around $N_e$ e-folds of inflation given by 
\be N_e \sim \f{H_c^2}{m^2} \sim \f{\rho_c}{m^2} = \f{m}{m^2}= m^{-1} \ee
e-folds of inflation wherein we denote the time at which the field becomes classical with a subscript $c$. This is approximately $10^6$ e-folds for the standard mass inflaton, and those minimally expanding undergo on the order of a single e-fold. We therefore see that however parameters are distributed by eternal inflation, this initial distribution must be incredibly heavily re-weighted by classical inflationary considerations before observations are made. Repeating the above reasoning with an inflaton potential given by $V=\lambda \phi^4$ gives a similar result, but in this case the number of e-folds follows $\lambda^{-1/3}$, with eternal inflation ending when $\rho<\lambda^{1/3}$. 

To take this further illustrate this, let us consider the background model without perturbations. Again for simplicity of exposition, we will assume that the potential for the the inflaton is a quadratic, and that the field value of the inflaton is distributed at some high initial energy density. Using the slow roll approximation, we find that the number of e-folds is approximately given by the square of the initial field value of the inflaton. Then we consider observing some physical parameter at the end of inflation (some lower energy density), again characterised by the one free parameter - the field value of the inflaton. Thus the probability of making an observation $o$ in some range $O$ relates to a corresponding set of compatible values of the inflaton field: As such, we find that an initial distribution translates to a final distribution by re-weighting:

\be P(o \in O) = P(\phi_f \in R) =\f{1}{N} \int_R d\phi_f = \f{1}{N'} \int_A \Omega \approx \f{1}{N'} \int_{R'} \exp[-\phi_i^2] d\phi_i \ee

in the final step, we have made use of the re-weighting via Liouville's theorem, and the slow-roll approximation that the number of e-folds is given by $\phi_i^2$. We have set $A$ to denote the thickened interval over which the symplectic form is integrated, and $R'$ is the corresponding range to $R$ at the higher, initial energy density. Thus, if we believe that after eternal inflation, some physical parameters are spread over a region of field values, we have had to significantly re-weight them in making our evaluation of the probability of making an observation at low energy density. Assuming a uniform distribution at low energy densities is the equivalent of a strong re-weighting at high energy densities, and the inverse holds: To correctly calculate low density probabilities we must re-weight high energy distributions, and this weighting can be very strong. 

\section{Conclusions} \label{Conclusions}

The notion of ``probability'' and associated terms such as ``generality" in a cosmological context is fraught with philosophical issues \cite{Barrow:2015fga,Wald,Freivogel:2011eg,Davies:2004ep}: It is unclear how one can use a frequentist approach when addressing a phenomenon which happens only once, and Bayesian methods require a prior, the choice of which can be highly influential on results. It is therefore valuable to examine generic behaviours which avoid such dependence on external parameters. A probability space consists of a triple: A sample space $\mathcal{S}$, a set of events $\mathcal{E}$ consisting of subsets of $\mathcal{S}$, and a measure assigning a probability to each event. In the cosmological context, we choose our sample space $\mathcal{S}$ to be the space of physically distinct trajectories, and can individuate these on a Cauchy surface determined by constant mean curvature (i.e. $H=const$ slices). An event can therefore be determined to be the set of trajectories for which a phenomenon occurs, such as those for which galaxies form. A probability measure will consist of a volume form on the space of physical observables, and we thus determine the probability of an event to be the integral of this measure over the range of physical observables on a given Cauchy slice determining those trajectories for which the event occurs, subject to the Hamiltonian constraint. These physical observables are taken to be (functions of) field positions and their momenta, which we have split into background/homogeneous variables and a (finite number) of Fourier modes, and thus these form a subspace of the system's phase space, with orthogonal directions being gauge. What we have shown above is that any measure on such variables can be ``thickened" to a volume form on the intersection of phase space and the constraint and Cauchy surfaces. Since volume forms are unique up to a choice of function any probability measure can be expressed as a function of physical variables multiplied by the Liouville measure, integrated over a finite range in the gauge directions.  This set of trajectories, as measured by the Liouville measure, is invariant under evolution, and therefore at a later Cauchy slice, a new function of physical observables can be induced such that the probabilities assigned to events is equal.  This function can be formed by evolving the range of physical observables determining an event from the initial to final slices. 

In adopting a relational interpretation, we should expect there to be gauge directions in our theory space due to the freedom of unit choice, which are used to define dimensionful quantities. For many gauge directions this action would be independent of the Cauchy slice chosen. The interval over which the measure has been evaluated could shift or expand so long as it was done equally across physical observables, and a probability obtained by counting the same set of trajectories with a measure at a later slice would be unchanged. However in the case of volume as we have shown there is a trajectory dependent expansion in the gauge direction. Thus, to match a probability distribution at some initial Cauchy slice, any distribution at a later slice must be weighted by the relative expansion in the gauge direction - i.e. the change in volume between slices. This relative weighting can be extremely large: the maximum number of e-foldings in the standard $m^2 \phi^2/2$ inflationary scenario is approximately $3H^2/m^2$ from an initial Hubble $H$ to the end of inflation. The minimum is around 1, so for observationally preferred values of mass and starting at the Planck scale the relative weighting reaches $\exp[10^{12}]$ \cite{DaveThesis}. It would, therefore, take a measure which has an extremely large preference for low expansion rates to overcome this. Volume weighting physical observables is not an artificial process, but rather one that is naturally induced by the evolution of dynamical trajectories.

We are then lead to consider the following: Suppose that the state of physical observables is determined at some initial high energy density according to a probability distribution - matter is spawned closed to the Planck density in a state determined by some unknown probability distribution. To recover these initial probabilities, any probability distribution on physical observables at later times must be weighted by the volume expansion between this initial energy density and the point at which observations are made. These higher weightings correspond to trajectories which experienced long periods in which the equation of state of the matter content was approximately $P=-\rho$, and the weightings are very strong. It would therefore be very unlikely that we would observe a universe which had not undergone such an expansion. The most heavily weighted universes are close to isotropic, and have had little energy in their inhomogeneous modes, with the bulk energy density being made up by a homogeneous scalar field with $w=-1$.

\section*{Acknowledgments}

This publication was made possible through the support of a grant from the John Templeton Foundation. The opinions expressed in this publication are those of the authors and do not necessarily reflect the views of the John Templeton Foundation. The author is indebted to Hans Winther for pointing out a number of typos, and to the editor and anonymous referee whose comments have improved the manuscript. 

\bibliographystyle{kp}
\bibliography{FullArticle}

\end{document}